\documentstyle[12pt,aaspp4]{article}
\begin{document}

\title{HIGH VELOCITY LINE EMISSION IN THE NLR OF NGC 4151\altaffilmark{1}}

\author{J.B. Hutchings}
 \affil{Dominion Astrophysical Observatory,
National Research Council of Canada,\\ Victoria, B.C. V8X 4M6, Canada}

\author{D.M.Crenshaw\altaffilmark{2}, A.C.Danks\altaffilmark{3}, 
T.R.Gull, S.B.Kraemer\altaffilmark{2}}
\affil{NASA's Goddard Space Flight Center, Lab for Astronomy and Solar
Physics, Code 681, Greenbelt MD 20771}

\author{C.H. Nelson, D. Weistrop}
\affil {Dept of Physics, University of Nevada, Las Vegas,
4505 Maryland Pkwy, Las Vegas, NV 89154-4002}

\author{M.E.Kaiser}
\affil{Dept of Physics and Astronomy, Johns Hopkins University, 
Baltimore, MD 21218}

\author{C.L.Joseph}
\affil{Dept of Physics and astronomy, Rutgers University,
Piscataway, NJ 08855-0849}

\altaffiltext{1}{Based on observations with the NASA/ESA \it Hubble Space 
Telescope, \rm obtained at the Space Telescope Science Institute, which 
is operated by AURA Inc under NASA contract NAS5-26555}
\altaffiltext{2}{Dept of Physics, Catholic University of America, Washington DC 20064}
\altaffiltext{3}{Raytheon ITSS}

\begin{abstract}
 Narrow-band imaging of the nuclear region of NGC 4151 with the Hubble Space
Telescope is presented. The filter bandpasses isolate line emission in
various high velocity ranges in several ions. Slitless and long-slit spectra
of the region with the Space Telescope Imaging Spectrograph also indicate
the locations of high velocity gas. These emission regions are faint and
are interspersed among the bright emission clouds seen in direct images.
They have radial velocities up to 1400 km s$^{-1}$ relative to the nucleus, 
and are found in both approach 
and recession on both sides of the nucleus. This contrasts strongly with the
bright emission line clouds which have been discussed previously as showing
bidirectional outflow with velocities within 400 km s$^{-1}$ of the nucleus. We
discuss the possible connections of the high velocity material with the radio
jet and the nuclear radiation.

\end{abstract}


\section{Introduction}

The Seyfert galaxy NGC 4151 is close enough and well oriented to enable a 
detailed study of the associated cloud structure around the nuclear engine, 
with the ultimate goal of understanding the Seyfert mechanism. HST 
instrumentation has advanced these studies with its high spatial imaging 
(e.g. Evans et al 1993, Boksenberg et al 1995, Winge et al 1997). The
power of HST was greatly extended by the installation of the Space Telescope
Imaging Spectrograph (STIS), with both slitless and longslit capability. 
The structure and kinematics of the narrow line region of NGC 4151 have been
studied recently using slitless spectra obtained with STIS (Hutchings 
\it et al \rm, 1998, Kaiser \it et al \rm, 1999). More detailed investigation
along subsequently defined long slit locations have been done by Nelson 
\it el al \rm (1999) and Kraemer \it et al \rm (1999). This work revealed a much
fuller picture of the kinematics and ionization structure of the ensemble of
emission line clouds that form a "biconical" region on both sides of the 
nucleus and roughly along the axis of the extended radio structure on either
side of the nucleus. The majority of the clouds appear to be outflowing from 
the nucleus at velocities $<$400km. s$^{-1}$ , approaching us to the west 
side and receding on the 
eastern side. These clouds appear to be part of a general outflow away from 
the nucleus, but they are not obviously connected with the radio jet outflow.
Beyond the inner 5 arcsec, the emission line clouds have lower velocities and 
appear to be participating in the general galaxy rotation rather than under 
any influence of the nuclear processes. They do appear to be illuminated 
and ionized by the beamed nuclear radiation.

  In the course of the work on the bright clouds, some weaker and highly
shifted emission components were seen in the slitless spectra. Kaiser 
\it et al \rm
(1999) identified 6 of these with velocities ranging from +846 to -1716
km s$^{-1}$. The location of the material with
these high velicites was established by use of two different dispersion
directions in the slitless data, as described by Kaiser \it et al \rm. These
locations did not generally correspond with any bright region seen in the
undispersed narrow band images, but lay within the ensemble of such clouds,
which have much larger flux than the high velocity material, as measured from 
the dispersed images.

   The high velocity material did not fit the simple biconical outflow 
scenario that describes the bright clouds, and Kaiser \it et al \rm noted 
that there seem to be more than one
kinematic components superposed in the region. In addition,
there is a bimodal distribution of velocity dispersions among the bright
clouds that is not related to the radial velocity, although the
high velocity clouds all have high velocity dispersion.

   In order to study the high velocity material in more detail, and to
distinguish it from the bright clouds, we obtained a number of narrow band
filter images with STIS and WFPC2. These have bandpasses near to the 
wavelengths of several strong emission lines, and comparison between them 
was intended to indicate the location of emission material in various 
velocity ranges. In all cases, these ranges are more than 400 km s$^{-1}$ 
from that of the nucleus, and thus help isolate the high velocity gas.

   We report here on these imaging data, and use them in conjunction with
the slitless spectra, and two long-slit STIS spectra through the region, to
discuss the high velocity gas in the nuclear region of NGC 4151.

\section{Image data}

    Table 1 shows the observational log. The redshift of NGC 4151 is close
to 1000 km s$^{-1}$, and the high velocities range over more than 2000 km s$^{-1}$
about that, so that the bandpasses of the narrow-band filters do not in general
include all the velocities present in the region. Figure 1 illustrates
the velocity ranges that are covered for the lines in the passbands,
as defined by the FWHM of their transmission functions. In
several cases there are two or more lines within the passband, so that
different velocities are sampled for the different lines. These are
the [O II] 3727A doublet and the [S II] 6717-31A doublet, and the [N II] plus
H$\alpha$ triplet at $\sim$6548-84A. We also obtained continuum images and
used an archival F502N image from WFPC2.

   Figure 2 shows some images and differences of interest. The LRF image
of [O III] covers all known velocities from the slitless data, and thus 
is the full representation of the [O III] flux. The STIS [O III] image
covers velocities between -1200 and -860 km s$^{-1}$ with respect to the nucleus.
It shows almost no resemblance to the LRF image, and some of the bright knots
are in areas of low flux in the total intensity image. Thus, the high
velocity material in this velocity range is much fainter than the main bright
clouds. It also is seen on both sides of the nucleus and outside the
main biconical emission region. The difference between the F502N and
LRF image shows regions of high positive velocity. These too do not correspond
with the main bright knots, and lie on the E side of the nucleus.

   The other difference images in Figure 2 are more complex as they compare different lines and combinations of lines. They do however indicate the same
regions of high positive and negative velocity, although the contrast is 
lower because of including continuum or some flux from the low velocity lines.

\section{Spectroscopic data}

    The slitless spectra have been described fully in the Hutchings \it et al \rm and
Kaiser \it et al \rm references. In Figure 3 we show one of the [O III] 5007A line,
intended to show the high velovity material clearly.
The locations of the high velocity material in the total [O III] image are
ahown in Figure 4. The long slit data are 
described in detail by Nelson \it et al \rm (1999). For the purpose of this 
paper we show the orientation of the two 0.1" wide slits in Figure 5.

    From the slitless data we can locate the high velocity regions labelled 
with letters in Figures 4 and 5. These positions depend on identifying
the same feature in both dispersed images. In crowded regions there is
occasionally some ambiguity between regions that lie along the same
dispersion direction. The image data have been valuable in showing up regions
where some high velocity material lies, and provide significant overlap
with the slitless spectra. 

The two long slit spectra do not
sample the whole region, but they do reveal a few high velocity regions that
correspond with (parts of) H, D, N(=26), and M, as well as verifying the 
bright knot velocities reported in the slitless papers (see Figure 5).
Table 2 shows the velocities from all our spectroscopic data. These differ
from values given for some of them by Kaiser \it et al \rm (1999), mainly because
of more detailed attention to these weak features. The differences are
in most cases within the uncertainties, and in most cases the large
line width (often blended with other stronger features) make the values 
reliable only within $\sim$10\% of their given value.  

The long slit data do not show
any signs of E, F, B, C, I, or J, although their estimated positions lie 
close to the slits. The positions for H and D also lie near the slit edges,
but so do G and A, which have similar velocities. It is thus possible that
there is extended high velocity emission lying between G-H and A-D. Since there
is no doubt of the reality of the high velocity material at  E, F, B, C, I, 
and J in the slitless images, we conclude that either the regions are compact,
or faintness and uncertainty in their locations explains why they
are not seen in the long slit data.

   The long slit data do indicate material with high positive velocity
(+400 km s$^{-1}$) near bright cloud 19 of Hutchings et al (1998), 
and some material at $\sim-$600 km s$^{-1}$
inside region M. The STIS [O III] image also indicates
regions of high negative velocity marked with small squares in Figure 4.
These are not seen in the slitless data as these velocity shifts place the
dispersed line emission in the region dominated by the bright emission 
clouds. Thus, it is clear that we do not have the complete picture of high
velocity emissions, and that a full set of long slit spectra, possibly 
with different orientations, would be required to get that.

   However, with the exception of bright knot 26, the high velocity
line emission has low flux and comes from locations superposed on the
whole inner bright emission cloud region. Region O also lies at the end
of a plume of material seen extending out of the main emission cloud
region, although very close to the nucleus.

  Winge \it et al \rm (1999) report velocities from several long slit spectra
obtained with the FOC. These have lower resolution and S/N and greater
distortions than the STIS spectra. However, their 
results also show up high velocity material that stands apart from the
lower velocities found in the main cloud ensemble. Their results detect
the material we designate N and G-H in their nuclear spectrum; C and G-H in 
their 0.23" NW offset; D and the knots seen in the STIS [O III] image in their
0.41" and 0.46" offsets to the NW. Thus their results are consistent with
those reported here. 

\section{Ionisation structure}

   With the complete wavelength coverage of the [O II], [O III], and H$\beta$
line velocities in our images, it is possible to construct ratio images 
that show the variation of O ionisation and [O III]/H$\beta$ in the 
emission line region. Figure 6 shows these ratios. The [O III]/[O II]
ratio was not corrected for the different continuum and nuclear scattering
in the two images, but these appear to be small as there is no 
systematic gradient away
from the nucleus. Both ratio images were tested empirically for
sensitivity to alignment, sampling, and rotation and the morphological
features that are seen are robust to any expected uncertainties.

    We first note that the ionisation of O is higher along radial locations
on both sides of the nucleus. The rays are longer to the W, which is the
approaching side for the general outflow. This suggests these are paths
where the nuclear radiation is less obscured, and ionise the gas more
strongly. These paths are seen more clearly in the approaching cone. 
Making a detailed comparison with the high velocity gas locations, we find
that many of them lie in the regions of higher ionisation.
Some (e.g. G and H, B, C, I, J lie around the edges of
a high ionisation region. A few (E, F, and A) lie in regions of low or
average ionisation. Region O lies in the part of the diagram affected by
the telescope diffraction spikes, so we cannot comment on it.
Generally there is an association of high velocity with higher ionisation.
There are, however regions of high ionisation (mainly to the E) with
no known high velocity gas. In view of the higher signal from the bright
clouds and the fact that not all high velocity regions appear to lie
in high ionisation places, we need to confirm these conclusions with systematic slit spectroscopy.

  Turning to the [O III]/H$\beta$ ratio, there is good signal over a smaller
region, but the main radial features are still present. This presumably
is a result of the same cause: where O is more highly ionised, the H is
weaker because it is ionised. Kaiser \it et al \rm have discussed the quantitative
[O III]/H$\beta$ values for many bright clouds based on dispersed images.
The present images will superpose the high velocity fluxes on the bright
clouds, and the latter presumably dominate. In detail, the high velocity
gas locations lie in or close to regions of high [O III]/H$\beta$, even
more markedly than the O line ratio. Once again, we have a few regions
of high ratio that do not lie near any known high velocity gas, and the
same cabest apply to any general conclusions about the high velocity gas.

   Thus, the high velocity gas is weak in H$\beta$ and [O II], and appears
to occur in highly ionised material. In the bright clouds Kaiser \it et al \rm
find a decrease in [O III]/H$\beta$ with increasing nulcear distance.
The high velocity gas has higher [O III]/H$\beta$: it is not known how high 
but the non-detection in [O II] and H$\beta$ implies numbers larger
than 15. The measured bright cloud line ratios from Kaiser \it et al \rm do not
show up as features in the ratio image, so that there may be gradients within
large clouds, or confusion from overlap. Table 2 shows nuclear distances
for the high velocity regions, and there is no correlation of line ratio
with this: the regions all lie within the high ionisation inner region of
Figure 5.

\section{Radio structure}

   The radio structure of NGC4151 lies along a line with the brightest feature
at the nucleus. There are several bright knots along the structure, on both 
sides of the nucleus, but with more flux and structure on the E (approaching)
side. As noted in Kaiser \it et al \rm, there is little detailed correspondence
between the bright clouds and the radio structure, although the radio
axis lies along the overall inner emission line region. In addition, the
innermost radio structure shows a bend away from the large-scale axis by
some 50$^o$ (to the N), but on a spatial scale unresolved by HST.

   Figure 4 shows the radio structure ridge line and the locations of the
main bright features at 6cm. The brightest knot C3 lies close to the brightest
emission line cloud \#20 from Hutchings \it et al \rm. 
Several high velocity gas locations lie close
to the radio ridge line (E, F, D, A), while others occur near bright radio
knots or bends in the ridge line (G, C, I). About half the high velocity
locations do not have any apparent relationship to the radio structure.
There is no correlation between the magnitude of the gas velocity and
the projected distance from the radio jet. There is also no high velocity
material in the direction of the innermost (VLBI) radio jet orientation.

\section{Discussion}

    Table 2 summarizes what is known about the high velocity material.
The high velocity gas appears to be a different kinematic component of
the nuclear environment from the bright biconical clouds. The high velocity
gas has higher ionisation, probably due to less obscured illumination
by the nucleus. The material has much lower flux (by more than an order
of magnitude in [O III] flux, and even more in other lines) than the
bright clouds, with the single exception of cloud 26 whose flux is
typical of the bright clouds. The high velocity material is seen with
velocities of approach and recession on both sides of the nucleus, quite
distinct from the clear separation of approach and recession across the
nucleus, in the bright clouds.

   The location of the known high velocity material (which is probably not
complete) is along ridges of higher ionisation. The ridges
are more marked and longer on the W side, where the bright clouds are
approaching (i.e. the near side of the bicone). The central ridge lies along
the line of the radio jet axis, and the radio ridge line appears
to wind around regions of highest ionisation (Figure 4). This may indicate 
that the jet is deflected by higher density in these regions, or that it
adds to the ionisation by its interaction with gas clouds, as proposed in
a more general sense by Winge et al (1999).

   12 of the 17 known regions of high velocity material lie in the W
(approaching) cone. This implies that there is less overall obscuration on 
this side, considering that the flux from all such regions is low. There is
no flux difference across the nucleus, in the regions that are seen.
The high velocity material does appear to be distinct from the bicone
outflow inferred from the bright clouds, but its overall velocity
and visibility is consistent with that picture. 

   All of the high velocity material has a high velocity dispersion, by
the criterion of 130 km s$^{-1}$ set by Kaiser \it et al \rm. Some of the velocity dispersions
(as noted in the table) are much higher, but also not well determined
because of blending in the slitless images. However, there are also
high velocity dispersion bright clouds with low radial velocities, as 
shown by Kaiser \it et al \rm.  There are some 10 bright clouds that have velocity
dispersions comparable with the low flux high velocity regions. 
High velocity gas is found only within about 1 arcsec of the nucleus, and
all high velocity dispersion material lies within about 2 arcsec of the nucleus.

   It is tempting to suppose that the high velocity material is accelerated
radially away from the nucleus with a velocity gradient that is also
radial. That would predict that there is a correlation between radial 
velocity and velocity dispersion due to projection effects, and we do not
in fact see clear evidence of this. Also, it would not explain why there are
bright low velocity, high velocity dispersion clouds. Thus, it
appears that this effect is not a strong one, and high velocity dispersion 
is found in both populations - bright outflow clouds and the high velocity gas.

    The high velocity gas is generally lower in density and appears in
places that may be directly illuminated by nuclear radiation or shocked by
nuclear material. The concidence with radio features may indicate that the jet
also moves along one line of direct illumination (travel) from the nucleus,
rather than direct acceleration of the high velocity gas by the jet.
 
    The fact that velocities of approach and recession are not connected with
position E or W of the nucleus suggests that the whatever drives the high
velocity material may push it
aside from the general radial motion, so that it may
appear with radial velocity counter to the general flow, by projection.
It may be that there is high velocity material projected with low radial
velocity
that is not recognised against the backdrop of lower velocity motions of
the bright clouds. We also note that the high velocity material may be
consistent with the ballistic ejection model proposed by Smith (1994),
as it is faint, has large velocity dispersion, and may be highly ionised.

  Finally, we note that the high velocities are similar to those seen strongly
in absorption in C IV, against the nucleus. However, those absorbers have
very low velocity dispersions (Weymann \it et al \rm 1997), with one exception.
The exception is a broad transient absorber seen at high approach velocity
in one GHRS spectrum. The location of the high velocity emitting gas
suggests a low covering factor against the nucleus from our line of sight.
The relationship of the narrow high velocity C IV absorbers
to the optical high velocity emitters is not clear.

\newpage

\centerline{References}

Boksenberg A., \it et al \rm, 1995, ApJ, 440, 151

Evans I.N., Tsvetanov Z., Kriss G.A., Ford H.C., Caganoff S., Koratkar A.P.,
ApJ, 417, 82

Hutchings J.B., \it et al \rm 1998, ApJ, 492, L115, 1998

Kaiser M.E., \it et al \rm 1999, ApJ (in press: Astro-ph 9906283)

Kraemer S.B., Crenshaw D.M., Hutchings J.B., Gull T.R., Kaiser M.E.,
Nelson C.H., Weistrop D., 1999, preprint

Nelson C.H., Weistrop D., Hutchings J.B., Crenshaw D.M., Gull T.R., Kaiser M.E.,
Kraemer S.B., Lindler D., 1999, preprint

Pedlar A., \it et al \rm, 1993, MNRAS, 263, 471

Smith M.D., 1994, MNRAS, 209, 913

Weymann R.J., Morris S.L., Gray M.E., Hutchings J.B., 1997, ApJ, 483, 717

Winge C., Axon D.J., Macchetto F.D., Capetti A., 1997, ApJ, 487, L121

Winge C., Axon D.J., Macchetto F.D., Capetti A., Marconi A., 1999, 
Astro-ph 9901415

\clearpage

\centerline{Captions to Figures}

1. Ranges of radial velocity with respect to NGC 4151 nucleus covered by
narrow-band filters, for emission lines that lie within their passbands.
The filters are shown and exposures are listed in Table 1. The dots in the
bottom row indicate the approximate velocities of the known high velocity
emission lines in NGC 4151. The range between -400 and 400 is occupied 
by almost all the bright emission line clouds.
 
2. Images and difference images of the central $\sim$5 arcsec of NGC 4151, as
labelled. The orientation is the same for all and North is indicated in 
Figure 4. The central emission line region shown is approximately 1 x 4 arcsec.
The bright clouds are approaching on the lower side and receeding on the
upper side of the nucleus. The material visible has velocity ranges as shown 
in Figure 1. Individual images are discussed in the text.

3. A slitless spectral image of [O III] line showing dispersed locations of 
high velocity material. The lower panel is the unprocessed image with the
nuclear continuum removed, and stretch to show the faint high velocity
components. The upper lanel is an unsharp mask that shows some of the high
velocity features more clearly.  

4. Positions of high velocity material identified on the total [O III] image.
Table 2 gives details of the individual positions as labelled. The key shows
how the positions are derived. The jagged line sketches the ridge of the radio
structure, with its bright knots identified by X and numbered following
Pedlar \it et al \rm 1993. The arrow indicates North.

5. The positions of our long slits shown with respect to the lettered high
velocity material from Figure 3. The slit widths are somewhat oversized 
and relative positions of the high velocity gas are uncertain by up to 0.1".

6. Matched [O III] and ratio images of central NGC 4151. Note the radially
extended regions of higher ionisation, particularly to the W (lower right). 
The images are discussed in the text.

\begin{deluxetable}{llrll} 
\tablenum{1}
\tablecaption{NGC 4151 Journal of observations}
\tablehead{\colhead{Instrument} &\colhead{Filter} &\colhead{Exp (sec)}
&\colhead{Date} &\colhead{Dataset}}
\startdata
STIS       & [O II]    & 300    & Dec 1996  & o3vs01030\nl
            &[O III]   & 300    &  & o3vs01040\nl
WFPC2       &F502N     & 870     &Jan 1995   &v2i50101A \cr
           & F658N     & 402    & Apr 1998   &u4230601A\nl
            &F390N     &1200                &&u4230604A \cr
            &F673N     &1200                &&u4230606A \cr
            &F656N     & 700                &&u4230608A \cr
            &F547M     &  10     &Feb 1999   &u5780207-8R \cr
            &FR533N [O III]    & 140         &&u5780203-4R \cr
            &FR533N   H$\beta$  & 300        &&u5780205-6R \cr
\enddata
\end{deluxetable}

\vskip 20pt 

\begin{deluxetable}{lrrrll} 
\tablenum{2}
\tablecaption{NGC 4151 high velocity emission lines}
\tablehead{\colhead{Region} &\colhead{RV\tablenotemark{a}} &\colhead{Nuclear}
&\colhead{Radio ridge} &\colhead{Ionisation} &\colhead{Notes}\\
&\colhead{km s$^{-1}$} &\colhead{distance} &\colhead{distance}}
\startdata
       A &-1450 &1.7" &0" &Average \nl
       B &-1300 &1.6" &0.3" &Average \nl
       C & -900 &1.5" &0.15" &High \nl
       D &-1200 &1.5" &0" &Peak &In long slit\nl
       D' &400 &1.6" &0.1" &Average &Longslit only, at cloud 19 \nl
       E & -380 &1.8" &$<$0.1" &High H? &Max RV -500 \nl
       F & -440 &2.0" &0" &High H? \nl
       F'& -290 &2.0" &- &- &Co-located with F \nl
       G & 730  &1.5" &0.15" &Peak &Max RV 990, Near radio C5 \nl
       H & 650 &1.6" &0.2" &High H? &Max RV 820\nl
       I & 700 &1.7" &0.1" &High &Near radio bend \nl
       J & 650 &1.7" &0.3" &Average \nl
       M &-1150 &0.4" &0.15" &High &Close to nucleus \nl
     N=26& -760 &0.6" &0.15" &High &Bright cloud 26 \nl
       O & -870 &1.0" &0.5" &High? &Normal to jet, in STIS image\nl
      STIS &--\tablenotemark{b} &0.8" &0.2" &Peak &Only in STIS image \nl
      STIS &--\tablenotemark{b} &1.5" &0.2" &Peak &Only in STIS image \nl
      STIS &--\tablenotemark{b} &2.0" &0.4" &Peak &Only in STIS image \nl

\enddata
\tablenotetext{a}{With respect to nucleus}
\tablenotetext{b}{-860 to -1200}
\end{deluxetable}

\end{document}